# Understanding how the use of AI decision support tools affect critical thinking and over-reliance on technology by drug dispensers in Tanzania

Ally Salim Jr[1,2], Megan Allen[1,2], Kelvin Mariki[1,2], Kevin James Masoy[1,2], and Jafary Liana[3]

[1] Elsa Health Inc.
[2] Inspired Ideas Research Foundation
[3] Apotheker Consultancy (T) Limited

**Authors Note**

This document outlines the authors' experiences working with lower-level primary care providers in drug shops (Accredited Drug Dispensing Outlets) in Tanzania. This evaluation is a part of a larger research project called Afya-Tek, which is being implemented with support from multiple stakeholders and the Tanzanian government. Our teams appreciate all the support provided by all the stakeholders in venturing in these new directions.

Correspondence regarding this article should be addressed to Ally Salim, email: ally@elsa.health.

## Abstract

The use of AI in healthcare is designed to improve care delivery and augment the decisions of providers to enhance patient outcomes. When deployed in clinical settings, the interaction between providers and AI is a critical component for measuring and understanding the effectiveness of these digital tools on broader health outcomes. Even in cases where AI algorithms have high diagnostic accuracy, healthcare providers often still rely on their experience and sometimes "gut feeling" to make a final decision. Other times, providers rely unquestioningly on the outputs of the AI models, which leads to a concern about over-reliance on the technology. The purpose of this research was to understand how reliant drug shop dispensers were on AI-powered technologies when determining a differential diagnosis for a presented clinical case vignette. We explored how the drug dispensers responded to technology that is framed as "always correct" in an attempt to measure whether they begin to rely on it without any critical thought of their own. We found that dispensers relied on the AI's decision 25% of the time, even when the AI provided no explanation for its decision.

# Introduction

## The ADDO Program

Established in 2003, the Accredited Drug Dispensing Outlet (ADDO) program was created in response to identified gaps in quality and availability of essential medicines in rural and peri-urban regions of Tanzania (Benavides et al., 2013). The assessment that uncovered these gaps, which was conducted by the US non-profit Management Sciences for Health's (MSH) Strategies for Enhancing Access to Medicines (SEAM) Program, highlighted the challenges faced by millions of rural Tanzanians trying to access medications for common ailments like malaria and diarrhea. The assessment also highlighted that it was commonplace for untrained personnel working in drug shops, known as Duka la Dawa Baridi (DLDB), to regularly dispense potent antibiotics and other prescription medication (SEAM, 2005).

In response to these findings, which indicated poor management practices and significant unmet medical needs, the "Tanzania ADDO Program" was established through a collaboration between Management Sciences for Health (MSH) and the Tanzania Food and Drug Authority (TFDA, 2015). Since its introduction, the ADDO program has successfully overcome many of the challenges presented by SEAM, and the success of the program has led to widespread adoption and an impressive scale of more than 14,000 outlets accredited in the country (Pharmacy Council, n.d.). Other countries such as Nigeria, Ethiopia, Uganda, Bangladesh, and more, have also adopted the program for their own countries and contexts.

ADDOs contribute to the increased availability of primary care and essential medicines in the country, and play a significant role in the triaging of patients, who often present first at an ADDO before going to another health facility or laboratory (TFDA, 2015). Through the ADDO program, drug dispensers are trained to identify basic illnesses and can accurately identify conditions like pneumonia and malaria, the leading causes of child mortality in Tanzania (UNICEF, n.d.). The ADDO program improves the quality of dispensing services, increases the availability and quality of medications, and contributes significantly to the early diagnosis, referral, and treatment that is necessary to save lives and improve health outcomes overall (Benavides et al., 2013).

The business model for the ADDO program is that of a public-private partnership, where the individual drug stores are privately owned and independently operated by their owners.

Regulations and guidelines are determined by national health bodies such as the Tanzania Pharmacy Council and the Tanzania Food and Drug Authority (Benavides et al., 2013). This type of business model allows for the flexibility and benefits of a privately-owned business, such as the ability to maximize profit, while also receiving support and training from government bodies and international funders.

Despite the successes of the ADDO program, there have been some growing pains and challenges at scale as highlighted by other researchers and analysts (Dilip et al, 2015; Minzi and Manyilizu, 2013; Embrey et al, 2016) . An external analysis by Minzi and Manyilizu (2013) showed that the ADDOs were in fact not significantly different from the old DLDMs in the inappropriate dispensing of antibiotics and other prescription medication. The authors also highlighted the low literacy of basic pharmacology by the drug dispensers as one of the key causes of improper dispensing and misinformation. Another study by the World Bank (2016), showed that four out ten patients (4/10) are misdiagnosed at the primary care level, contributing even further to the misuse of medications at these first points of care.

## Decision Support Technologies for ADDOs

There are a number of ongoing initiatives to improve the ADDO program and address some of the challenges that have been identified: health training of drug dispensers has been improved to address knowledge gaps, business skills and record keeping has been added to training programs, drug dispensers have been offered commercial incentives such as loans, and evaluation and supervision by ward and district inspectors has been strengthened to improve drug shop quality (CHMI, n.d.). The SHOPS Plus program, funded by the United States Agency for International Development (USAID), also identified a number of ways that ADDOs could be maximized for service delivery, and they have been working with partners to pilot various interventions (SHOPS Plus, 2016). The private sector has also started investing in the drug shops, with small digital health companies providing tools for stock management and ordering.

One of the largest projects to invest in the ADDO program is Afya-Tek, a multi-phase project that was launched in the Pwani Region of Tanzania in 2019. Funded by Fondation Botnar, the project is working to improve the decision-making and quality of healthcare by providing digital health tools for improved linkages between various health actors, including ADDOs (Fondation Botnar, n.d.).

Within this set of solutions, there is a unique opportunity to leverage artificial intelligence (AI) and decision support technologies to augment the capacity of the ADDO dispenser and strengthen clinical decision making at the point of care. AI has shown to catalyze improvements in healthcare delivery and patient outcomes, and has been successful in other settings at supporting healthcare providers to make more accurate, evidence-based decisions for patients (Dreyer and Allen, 2018). Applied to the ADDO use case, these technologies have the potential to broaden the scope of knowledge of the dispensers, improve diagnostic accuracy, and decrease the inappropriate use of antibiotics.

## Reliance on Decision Support Tools for ADDOs

With the introduction of AI-powered tools in healthcare, there has been increasing interest in the topic of reliance (LaGrandeur, 2021; Kerasidou et al, 2022; Sutton et al, 2020). That is, how much a user depends on or trusts in an algorithm's outputs and recommendations. However, there is still a limited understanding of the factors that affect a clinician's reliance on decision support tools, as well as how reliance affects clinician decision making and relationship with their patients.

The issue of reliance is particularly interesting in the context of lower-cadre health providers who often have limited training and low-technology literacy. In the ADDO setting, if dispensers rely too much on the technology, they could miss common signs that might not be directly asked by the tool (a patient's gait, the smell of their breath, their temperment, etc), leading to reduced critical thinking and the blind acceptance of recommendations that are missing key information. Over-reliance on AI also has the potential to reduce interactions and conversation between providers and patients (Vayena et al, 2018), and to reduce the ability of clinicians to operate outside of decision support systems (Want et al, 2021, Sutton et al, 2020). On the other hand, if the dispensers do not rely on the technology at all, or are reluctant to take into account information provided by the algorithms, they could be missing out on evidence-based, alternative options that might not have been in their circle of knowledge. Under-reliance could also indicate that a tool is not useful to the dispenser, and the continued use of the technology could negatively impact their workflow and the relationship they have with clients.

There is a fine line between over-reliance and under-utilization, one whose position is still unknown to us. These tradeoffs are at the intersection of human-computer interaction, user experience planning, and technology performance and can help determine whether or not AI interventions are effective at supporting health decisions across multiple levels of care. Given the complexity and importance of even the smallest decisions made at the ADDOs, the extent to which the dispenser relies on assistive technology can have a large impact on patient outcomes and even the dispenser's business goals.

As a part of the Afya-Tek program, Inspired Ideas Research Foundation and Elsa Health Inc are studying the feasibility and impact of AI-powered health decision support tools in the ADDO setting. These tools support drug shop dispensers in identifying the likely causes of someone's illness and provide treatment and management recommendations based on national guidelines. In order to better understand the issues of reliance on assistive technologies, this research evaluated how reliant ADDO dispensers are on the responses provided by an AI-powered tool. The researchers sought to better understand what factors influence a dispenser's reliance on decision support tools, as well as how the perceived "correctness" of the AI's outputs affects the dispenser's confidence in the technology's decisions.

## Methods

ADDO dispensers (n=8) were recruited to participate in this study. These dispensers have been engaged with the research team over the course of a year to evaluate the implementation of AI-powered clinical decision support tools in their ADDOs. During in-depth interviews in November 2022, dispensers were introduced to this activity and provided informed consent. Participants were only told that they were testing out a new tool. Dispensers were given with a 10" tablet that included a different application from the tool being tested in the broader AI research.

The application included a set of 10 clinical case vignettes which were developed by medical doctors and selected specifically for this research (see Appendix 1). Each vignette was displayed on the screen and included all of the information needed to make a diagnosis. Participants were asked by the interviewer, a medical doctor, to go through each case vignette and determine the most likely cause of the patient's presentation. Participants selected one condition from a multiple choice selection of four possible conditions. All of the options were differential diagnoses for the correct condition and were therefore similar in presentation to the reference standard diagnosis of the case vignette.

After indicating their choice for the most likely condition, the application revealed the AI's decision (see Schema for AI's Selections). Participants were then given a chance to change their choice; they could either keep their original answer, change it to match what the tool recommended, or change to something else entirely. After submitting their final answer, the interviewer would then reveal the "correct" answer to the participant.

After completing all ten vignettes, the participant was asked a series of Likert-scale questions (1-5 scale) about the helpfulness, accuracy, and reliability of the AI technology. They were also asked demographic and technology use questions to gauge their technology literacy.

### Schema for AI's Selections

The AI used in the application was not actually making predictions or determinations. Instead, it was intentionally hard-coded to provide a random selection. For the purpose of this study, we will use the term "artificial intelligence" or "AI" to describe the technology since that is how it was portrayed to the participants.

The conditions for the AI's selections included:
- The AI would never select the same response as the participant's first response; its answer was always different.
- The AI would not provide any additional information on the reasoning it used to make the decision, just the choice it made.

For example:

| Differential Options | Participant Response | AI Tool Response |
| --- | --- | --- |
| Bronchitis, Pneumonia, Bronchiolitis, and Influenza | Influenza | Pneumonia |

To account for the randomness in the AI's selection, the interviewer provided the "correct" answer. This was not necessarily the correct differential diagnosis for the vignette, but instead is designed to either confirm the AI's response. The purpose of this design was to understand how the perception of the AI's accuracy affects the participant's reliance on the tool, not to evaluate the accuracy of the technology itself.

Data Analysis

Data was collected and stored on the tablet as the participants went through each clinical case vignette. At the end of the data collection, the set of data for all participants was downloaded and shared with the rest of the study team. The data collected was analyzed using basic descriptive statistics with the Nim programming language to identify means, perform counts, and get variances in the responses.

During the data collection process, the interviewer was also observing the participant's engagement with the tool. This qualitative data was recorded on paper and then transcribed and grouped by themes. Interesting topics or notes were used to enhance the study's outputs.

Results

Perceptions of the Technology

The entire group of ADDO dispensers that participated were female. All of the participants owned smartphones and reported regularly using the messaging app called Whatsapp. Only two of the participants reported that they had heard the term "AI", or its colloquial Swahili phrase "Akili bandia" (25%).

Additionally, the participants thought the AI used in the experiment performed well (mean = 4.37, variance = 0.484) and was reliable (mean = 4.5, variance = 0.5). Participants also reported that they believed the AI was helpful at supporting clinicians to identify the cause of a patient's illness (mean = 4.75, variance - 0.188).

Reliance on the Technology

In total, 80 clinical case vignettes were presented to the dispensers (10 per dispenser). Overall, it was found that dispensers changed their answer to match the AI's answer 25% of the time. 6.25% of the time they changed their answer to something else, but not what the AI said. 68.75% of the time, they left their answer unchanged. On average, it took about 2.3 vignettes for dispensers to first rely on the AI's decision. Table 1 shows the data for each ADDO dispenser.

ADDO dispensers accurately identified the reference standard diagnosis 31.25% of the time (prior to them seeing the AI's reponse).

|        | Vignettes 1 to 10 |    |    |    |    |    |    |    |    |    | U | C | CA |
|--------|---|----|---|----|---|----|---|----|----|----|------|-----|-----|
| ADDO 1 | U | CA | U | C  | U | C  | U | U  | U  | U  | 70%  | 20% | 10% |
| ADDO 2 | U | U  | U | U  | U | U  | U | U  | U  | U  | 100% | 0%  | 0%  |
| ADDO 3 | U | U  | CA| U  | U | U  | U | U  | CA | U  | 80%  | 0%  | 20% |
| ADDO 4 | CA| U  | U | U  | U | U  | U | U  | U  | U  | 90%  | 0%  | 10% |
| ADDO 5 | CA| C  | CA| CA | C | CA | U | CA | CA | CA | 10%  | 20% | 70% |
| ADDO 6 | U | U  | CA| U  | U | U  | CA| U  | U  | C  | 70%  | 10% | 20% |
| ADDO 7 | U | U  | CA| U  | U | U  | U | CA | U  | CA | 70%  | 0%  | 30% |
| ADDO 8 | U | U  | CA| CA | U | U  | U | U  | CA | CA | 60%  | 0%  | 40% |

*Table 1:* U: Unchanged - Dispenser did not change their answer to reflect the AI's answer, C: Changed - Dispenser changed their answer to something else, but not what the AI said, CA: Changed to AI - Dispensers changed their answer to reflect the recommendations of what the AI said.

## Discussion

Throughout our deployments of AI-powered decision support tools across various levels of healthcare, we have noticed two scenarios. First, clinicians using the tools sometimes pay little to no attention to the outputs and insights that the AI provides when making treatment or management decisions for their patients. We've noticed that although clinicians might internalize what the tool is suggesting, they still rely heavily on their own internal decision making process, even when it is incorrect. Alternatively, we have also seen clinicians regularly default to what the tool is suggesting, and in some cases even rationalizing why the AI's decision is correct over their own. We have also gotten comments from stakeholders that AI-powered decision support tools will make doctors complacent because they "chew all of the information for you and spit out an answer". These concerns have led us to thinking more about the issues of reliance on technology and how it affects health decisions, trust, clinician-patient interactions, and overall health outcomes.

In this study we found that one out of every four times (25%) dispensers changed their final answer after seeing the AI's response. We consider this to be very high, especially given that the AI was randomly selecting one of four possible conditions and provided no explanations to back up its decisions. We also observed that the dispensers were more likely to change their final answer for conditions they do not typically encounter at their drug shop (such as alcoholic ketoacidosis), and sexually transmitted infections (such as bacterial vaginosis), signaling low knowledge of certain conditions and low confidence in their first choice.

It is also interesting to note that none of the dispensers consistently relied on the AI after first changing their response to match the AI. This is contrary to our expectations, which were that if a dispenser started to see that the AI was correctly identifying the conditions over and over again, that they would, at some point, start to consistently rely on the tool. This could indicate that dispensers are critically thinking about their responses or that they are more familiar with some of the conditions shown at the end and therefore more confident. It is also possible that the setting of the study made dispensers feel like they needed to not always rely on the tool in order to "prove" themselves, even though the purpose of the experiment was not shared with them.

In addition, there was great variability in the reliance on the AI tool; one of the dispensers never changed their answer to match that of the AI, while another one changed theirs 80% of the time. A larger sample size is needed to further understand this variability, however from our observations there are likely personality traits (confidence, tenacity, self-efficacy), knowledge levels, and perceptions (technology is good/ useful, AI is helpful) that are affecting each individual's decision to rely on the technology.

Although not the main purpose of the study, we found that the ADDO dispenser's accuracy at identifying the reference standard diagnosis (the true answer for the clinical case vignette) across all conditions prior to seeing the AI's response was quite low (31.25%). This is especially true when compared to the random picker AI with an accuracy of 27.5%. This shows that there is a real need and opportunity for providing decision support to improve ADDO dispenser decisions. The combination of dispenser accuracy and reliance with a highly performant and user friendly decision support tool holds significant potential.

Lastly, explainability and interpretability of outputs plays a key role in the reliance on, and trust in, decision support tools. When explainability exists, clinicians are able to understand how and why AI reaches a certain recommendation, improving overall reasoning and collaboration with the tool. When explainability is missing or ineffective, clinicians either overly rely on the decisions made by the AI or ignore the insights entirely. There is no opportunity for a collaboration of the "minds" or the insights of a second opinion. In future iterations of this research, evaluating the difference between tools with explainability and those without could illuminate important findings.

Limitations

The biggest limitation is the sample size, which limits our ability to determine any significance in the findings. Although the small sample size was deliberate in order to begin exploring this space, expanding the number of participants will allow us to make more definitive conclusions. Another limitation is that the research activity was conducted in a "lab" setting as opposed to the real world. This could impact the ADDO dispenser's perception of the task and cause them to perform differently than they might if they were using the tool in their drug shop. Additionally, since the users were all part of the larger study evaluating AI tools in the ADDO setting, the tool tested in this activity could have been highly conflated with the tools they use in their drug shop. Any perceptions they have of the latter could have impacted how they participated in this research.

Although the structure of the activity was important for the type of information we wanted to collect, only allowing dispensers to choose from four conditions in a multiple choice format is also not fully indicative of how decision support tools typically function. Dispensers were not able to identify multiple possible conditions, nor were they able to select a condition that was not already coded into the platform. This could have impacted how they selected conditions.

Recommendations & Directions for the Future

It's clear that the implications of reliance on AI-powered decision support tools in the healthcare setting can have a significant impact on both the utility of the tools and the outcomes for patients. There is a need to better understand the various factors that influence a clinician's reliance on AI tools, as well as to explore further the effect of explainability and interface design on reliance and trust. This can be accomplished with larger sample sizes and varying levels of care providers. Another recommendation is to redesign the activity to collect data in a more "real-world" setting to better capture true reliance while the tools are being used.

Appendix 1

*Clinical Case Vignettes Used*

Vignettes were developed by a group of medical doctors from Inspired Ideas Research Foundation and Elsa Health. The conditions included in this study were presented in the following order:

1. Dengue
2. Emphysema
3. Alcoholic Ketoacidosis
4. Allergic Rhinitis
5. Bronchiolitis
6. Urethritis
7. Amoeba
8. Severe malaria
9. Bacterial vaginosis
10. Tonsillitis